%% file: main.tex
\documentclass[manuscript]{acmart}
\usepackage{makecell}
\AtBeginDocument{%
  }

\setcopyright{acmlicensed}
\copyrightyear{2018}
\acmYear{2018}
\acmDOI{XXXXXXX.XXXXXXX}
\acmISBN{978-1-4503-XXXX-X/18/06}

\begin{document}

\title[``We do use it, but not how hearing people think'']{``We do use it, but not how hearing people think'': How the Deaf and Hard of Hearing Community Uses Large Language Model Tools}

\author{Shuxu Huffman}
\email{shuxu.huffman@gallaudet.edu}
\affiliation{%
  \institution{Gallaudet University}
  \city{Washington}
  \state{DC}
  \country{USA}
}

\author{Si Chen}
\email{sichen34@nd.edu}
\affiliation{%
  \institution{University of Notre Dame}
  \city{Notre Dame}
  \state{In}
  \country{USA}
}

\author{Kelly Avery Mack}
\email{kmack3@uw.edu}
\affiliation{%
  \institution{University of Washington}
  \city{Seattle}
  \state{WA}
  \country{USA}
}

\author{Haotian Su}
\email{haotias@g.clemson.edu}
\affiliation{%
  \institution{Clemson University}
  \city{Clemson}
  \state{SC}
  \country{USA}
}

\author{Qi Wang}
\email{qi.wang@gallaudet.edu}
\affiliation{%
  \institution{Gallaudet University}
  \city{Washington}
  \state{DC}
  \country{USA}
}

\author{Raja Kushalnagar}
\email{raja.kushalnagar@gallaudet.edu}
\affiliation{%
  \institution{Gallaudet University}
  \city{Washington}
  \state{DC}
  \country{USA}
}

\renewcommand{\shortauthors}{Huffman et al.}

\begin{abstract}
    Generative AI tools, particularly those utilizing large language models (LLMs), are increasingly used in everyday contexts. While these tools enhance productivity and accessibility, little is known about how Deaf and Hard of Hearing (DHH) individuals engage with them or the challenges they face when using them. This paper presents a mixed-method study exploring how the DHH community uses Text AI tools like ChatGPT to reduce communication barriers and enhance information access. We surveyed 80 DHH participants and conducted interviews with 11 participants. Our findings reveal important benefits, such as eased communication and bridging Deaf and hearing cultures, alongside challenges like lack of American Sign Language (ASL) support and Deaf cultural understanding. We highlight unique usage patterns, propose inclusive design recommendations, and outline future research directions to improve Text AI accessibility for the DHH community.
\end{abstract}

\begin{CCSXML}
<ccs2012>
   <concept>
       <concept_id>10003120.10011738.10011773</concept_id>
       <concept_desc>Human-centered computing~Empirical studies in accessibility</concept_desc>
       <concept_significance>500</concept_significance>
       </concept>
 </ccs2012>
\end{CCSXML}

\ccsdesc[500]{Human-centered computing~Empirical studies in accessibility}
\keywords{Accessibility, LLM, ChatGPT, Deaf and Hard of hearing}


\maketitle
\pagestyle{plain}

\input{1-intro}
\input{2-relatedworks}
\input{3-methods}
\input{4-findings-RQ1}
\input{4-findings-RQ2}

\input{5-conclusion}

\bibliographystyle{ACM-Reference-Format}
\bibliography{sample-base}

\newpage
\appendix
\input{7-appendix}

\end{document}

%% file: 1-intro.tex
\section{Introduction}

Generative AI tools, such as OpenAI's ChatGPT and Microsoft Copilot, are widely used across professional and personal contexts, including email writing, programming, and accessibility support \cite{feuerriegel2024generative, laquintano2023introduction, morrison2023understanding, froehlich2024future}. Text-based generative AI, which we refer to as ``Text AI\footnote{We use this term because there is no standard ASL sign for ``LLM'' or ``Generative AI''. However, in the Deaf community people refer to such tools as MESSAGE-AI or TYPE-AI so we chose a more general term that is closest to the interpretation.}'', powered by LLMs, is particularly popular for its ability to process natural language inputs and generate human-like text \cite{feuerriegel2024generative}.

Text AI holds significant potential \cite{pack2023potential, feuerriegel2024generative} for the Deaf and Hard of Hearing (DHH) community, which is characterized by strong cultural identities and diverse linguistic backgrounds \cite{bat2000diversity, parasnis1998cultural, cannon2016increasing}. Many DHH individuals are bilingual, with sign language (e.g., ASL) as their primary language and written language (e.g., English) as their second language \cite{parasnis1998cultural}. Text AI can help bridge language gaps by improving fluency and clarity in written English, reducing communication barriers in a predominantly hearing society. However, the lack of research on how DHH individuals use Text AI limits understanding of its benefits and challenges. 
%Biases in AI systems further risk reinforcing existing accessibility barriers \cite{desai2024asl}.
This work addresses these gaps by exploring how the DHH community interacts with Text AI and identifying barriers to its use. Specifically, we investigate:

\begin{description}
    \item[RQ1] How does the DHH community interact with and utilize Text AI tools?
    \item[RQ2] What are the challenges and opportunities of using Text AI in the DHH community?
\end{description}

To answer these questions, we conducted a survey with 80 DHH participants and interviewed 11 DHH participants\footnote{Two interviewees also participated in the survey.} using ASL and English study materials. Our analysis reveals that DHH participants exhibit mixed feelings and opinions about their use of Text AI. On the one hand, they acknowledged meaningful benefits that AI tools offer in overcoming communication barriers in a hearing-majority society, including bridging Deaf and hearing cultures and increasing confidence in written English communication. On the other hand, they raised concerns about the tools' lack of DHH-friendly interaction techniques, education, guidance, and understanding of Deaf cultural nuances. Participants also offered valuable suggestions for improving Text AI tools to better meet the needs of DHH users.

%% file: 2-relatedworks.tex
\section{Background and Related Work}

\subsection{Background of The DHH Community} 

The Deaf\footnote{Capital ``D'' Deaf refers to individuals who identify with Deaf culture and use sign language, while deaf refers to the physical condition of hearing loss without necessarily identifying with the Deaf community\cite{bat2000diversity, parasnis1998cultural}.} community encompasses rich cultural and linguistic diversity, characterized by unique identities and experiences \cite{bat2000diversity, parasnis1998cultural, solvang2014accessibility}. Scholars advocate for a postcolonial framework over the traditional medical model to emphasize positive aspects of Deaf culture, including concepts like Deafhood, DeafSpace, and Deaf gain \cite{ladd2005deafhood, edwards2014deafspace, bauman2014deaf}. However, not all DHH individuals identify with Deaf culture, as connections to Deaf identity often vary due to linguistic and cultural factors \cite{cannon2016increasing}. While this paper focuses on the broader DHH community, attention is also given to the Deaf community's cultural and linguistic context.

Sign language, particularly ASL, is a cornerstone of Deaf culture and identity, recognized as a fully developed language with its own grammar and vocabulary \cite{stokoe2005sign, valli2000linguistics}. Features like ASL gloss--a notation system used to represent individual signs in written form \cite{wolbers2014description}--have proven effective in improving English literacy among Deaf children \cite{supalla2017american}. Additionally, DHH individuals adopt diverse strategies to communicate with hearing people, blending oral and visual methods \cite{domagala2019strategies}. Scholars highlight the need for culturally responsive research methodologies to meet the educational and cultural needs of DHH individuals \cite{cannon2016increasing, scott2021call}.

\subsection{Unique Challenges The DHH Community Faces}
The DHH community faces issues such as societal bias, discrimination, and ableism, many of which are experienced in other disability communities \cite{mcmahon2012overview, colella2013workplace, mack2024they}. However, there are also challenges that are specific to the DHH community.

\subsubsection{Difficulties in Language Acquisition}
Deaf children face significant challenges in acquiring spoken languages due to limited auditory input and reliance on visual communication, compounded by educational resources designed for hearing learners \cite{swisher1989language, kontra2017foreign, csizer2020foreign}. The historical emphasis on oralism over sign language use has resulted in language deprivation during critical developmental periods, leading to long-term academic, social, and emotional challenges \cite{hall2019deaf, hutchison2007oralism, hall2017you, glickman2018language}. Marginalizing ASL while prioritizing English without accessible instruction exacerbates difficulties in language acquisition, particularly with English fluency and grammar \cite{qi2012large, cannon2013grammar}. Researchers emphasize addressing these barriers through culturally informed educational approaches \cite{kontra2017foreign, rose2004language}.

\subsubsection{Discrimination and Trauma}
Discrimination against DHH individuals often stems from linguistic bias, misconceptions about sign language, and communication barriers, resulting in marginalization across workplaces, education, and social settings \cite{freeman1989cultural, juni2002direction}. Daily communication barriers can lead to frustration, trauma, and exclusion \cite{ma2022discrimination, das2024comes}. Audism, the belief in the superiority of hearing over deafness, further marginalizes DHH individuals, particularly sign language users, by reinforcing systemic inequalities in employment and social interactions \cite{o2022opportunity}.

The DHH community is also uniquely vulnerable to Information Deprivation Trauma (IDT), characterized by emotional distress caused by a lack of access to critical information \cite{schild2016information}. This is especially pronounced in a hearing-dominated society where spoken-language information is often inaccessible without alternatives such as sign language or captions. IDT disproportionately affects Deaf individuals, hindering their ability to understand, prepare for, or respond to significant situations \cite{schild2012trauma, schild2016information}.

This study explores the potential of Text AI, a language tool \cite{pack2023potential, feuerriegel2024generative}, to address these challenges. %By providing accessible language and communication support, Text AI could bridge gaps between Deaf and hearing cultures, improve access to information, and reduce emotional distress linked to communication barriers. While not a comprehensive solution, Text AI offers a promising avenue to enhance daily communication for Deaf individuals, addressing both its benefits and limitations.

\subsubsection{The DHH Community in the AI Era}
The rise of AI offers opportunities and challenges for the DHH community. While tools to improve English grammar exist \cite{cannon2011improving}, biases in AI—rooted in non-representative datasets and assumptions by hearing researchers—risk systematically excluding Deaf users \cite{desai2024systemic, gugenheimer2017impact}. Advances in tools such as automatic captions and sign language recognition and translation have improved accessibility in work and education \cite{arroyo2024customization, papastratis2021artificial, joze2018ms, camgoz2020sign}. However, little research explores how DHH individuals use generative AI tools. This study examines their usage patterns, benefits, challenges, and barriers with these tools.

%% file: 3-methods.tex
\section{Methods}

Our study consists of a survey and complementary interviews approved by the institution's IRB. The research team included Deaf and hearing members with varying signing abilities.

\subsection{Survey}

We conducted an online survey targeting DHH individuals. Participants met the following criteria: 18 years or older, U.S.-based, self-identified as DHH, and fluent in ASL, English, or both. Recruitment was conducted through social media, email lists, a Deaf non-profit organization, a university, and snowball sampling \cite{sadler2010recruitment}. To ensure accessibility, materials, including the consent form, were provided in both ASL and English \cite{mack2022anticipate}. Multiple-choice questions used straight-forward language, while open-ended questions allowed responses via text or uploaded ASL videos. At the survey's start, participants were given a definition of Text AI, explaining it as a program that generates text in response to user prompts, with examples like ChatGPT, Copilot, and Gemini. Participants were entered into a raffle to win one \$20 gift card for every 10 participants.

We received 90 responses, excluded 10 bot responses, and tabulated the remaining 80 valid responses. One participant opted for a video chat, and a Deaf researcher conducted the session. The average participant age was 37.5 years ($SD = 12.3$). Among respondents, 55 identified as Deaf, 13 as Hard of Hearing, 8 as deaf, 1 as DeafBlind, 1 as Deaf and Dumb, and 2 as other, specifying deaf/HoH\footnote{HoH refers to Hard of Hearing.} and Deaf/HoH. ASL was the primary communication language for 52 participants, 22 used English, and 3 used both depending on the situation. Additional languages reported included Armenian, Farsi, and Chinese.

The survey questions (provided in the Appendix) were inspired by our RQs and relevant literature \cite{mack2020social, heskett2001using}. The survey comprised six sections, structured around five key variables we aimed to analyze: 1. frequency of Text AI use, 2. types of tasks for which Text AI is used, 3. reduction in anxiety or perceived discrimination, 4. satisfaction with Text AI assistance, and 5. the difficulties and challenges encountered while using Text AI.

\subsection{Semi-structured Interview}

We conducted semi-structured interviews ($N = 11$) to explore how DHH individuals use Text AI tools in daily life. Among the participants, six self-identified as Deaf and five as Hard of Hearing. Participants, aged 20 to 42, represented diverse genders, ethnicities (Eritrean, Caucasian, Hispanic, Filipino, and Middle Eastern), and communication preferences, including ASL, other sign languages, English, or a combination, reflecting varied experiences within the DHH community. Recruitment was conducted through a university email list and expanded via snowball sampling \cite{sadler2010recruitment} with the same recruitment criteria as the survey.

Interviews included task-based interactions with ChatGPT, the most commonly used Text AI tool reflected in our survey. Participants completed three tasks: (1) drafting an email to a professor requesting a homework extension, (2) addressing a group teammate’s tardiness in an email, and (3) inquiring about an internship opportunity with a university recruiter. Since most participants were recruited from a university, these tasks were tailored to reflect common scenarios in university life. Follow-up questions assessed their satisfaction with ChatGPT's responses, typical usage patterns, and additional challenges or questions. Interviews were conducted over Zoom, lasting approximately 30 minutes, with shorter durations for some participants. To ensure accessibility \cite{mack2022anticipate}, ASL interpreters and Zoom’s auto-captioning were provided as needed. Participants were encouraged to ask questions and seek clarification at any time and received gift cards as compensation with an hourly rate of \$25.

\subsection{Interview and Survey Analysis}

%ASL interview responses were translated into English by a team member fluent in ASL. Both interview and survey data were analyzed using thematic analysis \cite{braun2006using, braun2019reflecting}, with two researchers independently developing codes, reconciling differences, and applying final codes through mutual verification and discussion. Key themes were collaboratively identified to represent the findings. Quantitative analysis on the survey data compared participants' feelings with and without using Text AI and examined the influence of primary language on their perceptions. While comparisons such as Deaf versus Hard of Hearing were possible, the analysis focused on language due to the data distribution and the nuanced spectrum of identities within the community, as confirmed by responses to open-ended identity questions.

A Deaf researcher translated the ASL interview responses into English. Two researchers conducted thematic analysis \cite{braun2006using, braun2019reflecting} by independently developing codes, reconciling differences, and mutually verifying and discussing the final codes. Together, they collaboratively identified key themes to represent the findings. For the survey data, quantitative analysis examined participants' feelings with and without using Text AI and explored how their primary language influenced perceptions. While comparisons such as Deaf versus Hard of Hearing were possible, the analysis focused on primary language (ASL or English) due to the data distribution and the nuanced spectrum of identities within the community, as reflected in responses to open-ended identity questions.

%% file: 4-findings-RQ1.tex
\section{Results and Discussion}

We first present usage patterns and experiences with Text AI systems, focusing on their role in  enhancing communication confidence and bridging cultural gaps (RQ1). We also address concerns such as loss of cultural nuances and suggestions for improvement (RQ2). Our analysis integrates insights from open-ended survey responses, interviews, and statistical data, with interview participants labeled as ``I\#'' and survey respondents as ``S\#.''

\subsection{RQ1-How Does The DHH Community Interact With and Utilize Text AI Tools?}

%Beyond common Text AI tasks within the general population (help writing emails, looking up facts), we identified several themes regarding how DHH individuals specifically interact with and use Text AI tools.
We identified distinct themes highlighting how DHH individuals uniquely interact with and utilize Text AI tools, presenting qualitative results first, followed by quantitative findings.

\subsubsection{Using Text AI to Enhance English Confidence and Language Learning}

Text AI tools help reduce language barriers for the DHH community, especially when communicating with English-speaking hearing individuals. Participants noted that using non-standard English, like ASL gloss, often leads to misunderstandings, criticism, and discrimination, affecting confidence \cite{freeman1989cultural}. For example, S2 shared, \textit{``Deaf people write in `broken' English... hearing people struggle to understand, but ASL users pick it up easily because it's written in ASL gloss.''} S79 also shared their experience in school: \textit{``I always forgot to add the `-s', `-ing', commas... teachers were always mad at me, and I lost my confidence. I never finished college English.''}

Participants reported that Text AI boosted their confidence in writing by providing nonjudgmental corrections and improving grammar, vocabulary, tone, and fluency. As S78 expressed, \textit{``It makes using English easier and better than before... it boosts my confidence and makes me feel more like myself.''} Another participant (S79) valued Text AI over a tutor in language learning, saying, \textit{``I’m not embarrassed. I just get my English corrected.''}

\begin{figure}[ht]
    \centering  
    \includegraphics[width=0.85\textwidth]{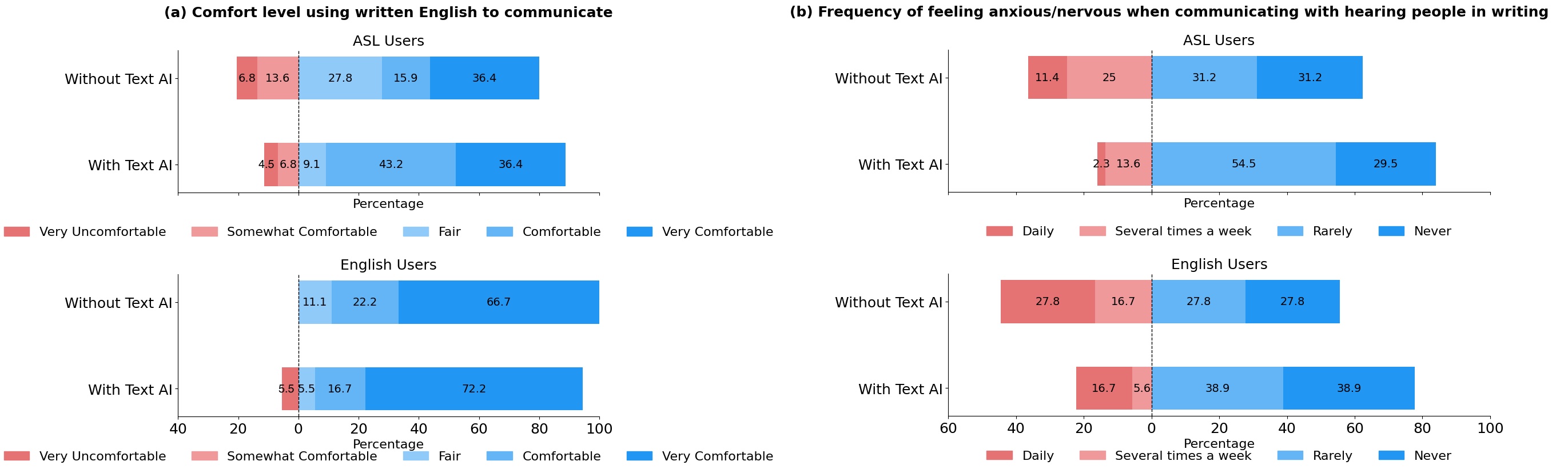}    
    \caption{The diagrams show data from Text AI users with ASL or English as their primary language (N = 62), comparing comfort levels and anxiety frequencies in written communication with and without Text AI. Data from 12 non-users, 3 participants with other primary languages, and 3 participants who identified both ASL and English as their primary languages are excluded. Diagram (a) shows changes in comfort levels using written English for ASL users (N = 44) and English users (N = 18). Diagram (b) presents changes in anxiety frequency when communicating with hearing individuals in writing. Text AI increased comfort levels for ASL users but slightly decreased them for English users in (a), while reducing anxiety frequencies for both groups in (b).}
    \Description{This figure displays two grouped bar charts comparing people's comfort levels and anxiety when communicating in written English, with and without the assistance of Text AI. The figures illustrate that the tool can assist with written English communication, and the diagrams provide a visual comparison before and after its use.

    Left Diagram (a):
    Title: ``Comfort level using written English to communicate.''
    The diagram compares individuals' comfort levels using written English without Text AI (upper) and with Text AI (lower), for both ASL users (top diagram) and English users (bottom diagram). Comfort levels range from ``Very Uncomfortable'' on the left to ``Very Comfortable'' on the right, represented by colors: red (Very Comfortable), pink (Comfortable), light blue (Fair), blue (Somewhat Uncomfortable), and dark blue (Very Uncomfortable). It shows a slight increase in comfort levels for ASL users with Text AI, while English users experience a slight decrease.
    
    Right Diagram (b):
    Title: ``Frequency of feeling anxious/nervous when communicating with hearing people in writing.''
    This diagram illustrates the frequency of anxiety or nervousness experienced during written communication with hearing people, comparing scenarios without Text AI (upper) and with Text AI (lower). Frequencies range from ``Daily'' on the left to ``Never'' on the right, represented by colors: red (Daily), pink (Several times a week), blue (Rarely), and dark blue (Never). The diagram shows a decrease in anxiety frequency for both ASL and English users when using Text AI.

    The figures indicate that the use of Text AI generally enhances comfort levels and reduces anxiety or nervousness in written English communication for many users, while English users may benefit less from it.}
    \label{fig:bar_comfort_anxiety}
\end{figure}

Survey data supports these findings. Among the 68 users with experiences of using Text AI, 55.9\% reported Text AI improved fluency, 50.0\% organization, 48.5\% vocabulary, 42.6\% grammar, 42.6\% tone and voice, and 45.6\% overall confidence. Satisfaction levels varied, with 58.8\% feeling satisfied or very satisfied with its assistance in improving English proficiency. Although the Wilcoxon Signed-Rank Test revealed no statistically significant difference in comfort levels before and after using Text AI ($p = 0.08$), we observed trends suggesting potential improvements. Without Text AI, 63.2\% of users felt comfortable or very comfortable using written English to communicate, compared to 82.4\% with Text AI, while those feeling somewhat uncomfortable or very uncomfortable dropped from 14.7\% to 8.8\%. Although most users experienced benefits, a few reported decreased comfort (Fig. \ref{fig:bar_comfort_anxiety}(a)), which we explore in later sections.

\subsubsection{Bridging Cultural Gaps and Easing Communication With Hearing People}

The Deaf community faces unique cultural challenges beyond language barriers, including communication needs and navigating differences with hearing culture. Our study found that Text AI can serve as a cultural mediator, bridging gaps between Deaf and hearing cultures in contexts ranging from everyday interactions to addressing ableism, and alleviating communication stress when interacting with hearing individuals.

One participant (S76) shared how ChatGPT helped bridge communication gaps with their hearing family, who often misunderstood their Deaf perspective: \textit{``My Deaf perspective and opinions just don't resonate with my family.''} Living between Deaf and hearing cultures, S76 found that carefully wording their messages reduced conflict but described this effort as \textit{``exhausting as heck.''} They spent 15 years away from their family due to communication challenges. After returning, they began using ChatGPT to reword their messages, requesting it to be \textit{``casual but sound firm.''} They explained, \textit{``Especially when my emotions are high, I type out what I need to express to ChatGPT (unfiltered), and it filters it in a way that is more acceptable to express.''} This tool became a vital intermediary, with the participant noting, \textit{``Somehow, every single time, ChatGPT wins. I use what it suggests, and things go more smoothly... And I feel closer to my family. There is less conflict.''}

Participants highlighted cultural differences in communication styles as a challenge, particularly the perceived bluntness of Deaf culture compared to the more indirect norms of hearing culture \cite{goss2003hearing}. Text AI tools were valued by our participants (I7, S59, S75, S22) for helping to adjust tone and reduce misunderstandings in these interactions. These tools also supported participants in navigating sensitive conversations about access and ableism. One participant (S44) explained, \textit{``I use ChatGPT to make my emails come off as more `polite' when I need to address access issues or ableism with my coworkers and supervisors.''} Text AI tools helped reduce the emotional burden of advocating for accommodations and addressing discrimination, which is often exhausting for DHH individuals.

Our quantitative data also indicates that Text AI significantly reduced communication anxiety experienced by DHH individuals when interacting with hearing people (Fig. \ref{fig:bar_comfort_anxiety}(b)). The proportion of users experiencing daily or weekly anxiety decreased from 14.7\% without Text AI to 5.9\% with its assistance, while those reporting never or rarely feeling anxious increased from 63.2\% to 83.8\%. The Stuart Maxwell test confirmed that the anxiety level reduction was statistically significant ($p < .01$). %The reduction of anxiety level varied across participants with different language preferences. Stuart Maxwell test results suggested that participants who primarily use ASL experienced a significant reduction in anxiety level ($p < .05$). In contrast, no significant change in the anxiety level was identified with participants who primarily use English ($p = .09$). 
This finding demonstrates Text AI's effectiveness to ease cross-cultural communication.

\subsubsection{Fact-Checking and Simplifying Information Retrieval} 

While Text AI tools face criticism for errors and aberrations \cite{feuerriegel2024generative}, participants highlighted their unique benefits for DHH users, particularly in clarifying ambiguous queries and reducing cognitive load during the process of gathering information. Unlike Google, Text AI's conversational format allows users to refine their language, making information retrieval more accessible for some DHH individuals. For instance, S79 described using one Text AI tool to clarify their query in English and another to seek information: \textit{``I get my thoughts together, use ChatGPT to make my English, then give to Gemini and ask `true or false?' ''} %This process supports fact-checking and helps users overcomes language-related challenges.

Summarization was another key feature for participants, enabling them to condense dense texts and meeting transcripts. As S8 shared, \textit{``I give a complicated wall of text and ask it to simplify for me,''} reducing the mental fatigue common among DHH individuals when processing complex content. Participants also used Text AI to summarize meeting transcripts, mitigating issues like rapid speech of multiple parties and captioning errors \cite{mcdonnell2021social, mcdonnell2023easier}. Text AI may further alleviate Information Deprivation Trauma (IDT) by providing accessible tools for forming questions, verifying facts, simplifying information, and bridging gaps in critical knowledge. However, its effectiveness relies on accuracy and cultural sensitivity, emphasizing the need for further refinements.

\subsubsection{Reduced Perceived Benefits of Text AI for Proficient English Users} 

Figure \ref{fig:bar_comfort_anxiety}(a) shows that most participants' comfort levels either increased or remained unchanged with Text AI, though some experienced a decrease, particularly those already comfortable with written English. While some participants appreciated that Text AI made their English more aligned with their intentions, others found the responses unnatural. This contradiction may stem from the DHH community's diverse cultural and linguistic backgrounds. Participants who described AI responses as ``fake'' (I14) or ``unnatural'' (I13) were predominantly college-educated, which suggests that higher English proficiency might increase sensitivity to linguistic nuances. Likert-scale data on a 5-point scale indicates that mean comfort levels for ASL users increased from 3.6 to 4.0 with Text AI, while English users experienced a slight decrease from 4.6 to 4.5. Participants with high English proficiency raised a concern for the broader DHH community: Text AI may lead DHH users with lower English proficiency to automatically trust its outputs, leading to misinformation, a trend also observed in Automatic Speech Recognition (ASR) technologies \cite{berke2018methods}.

%% file: 4-findings-RQ2.tex
\subsection{RQ2-What Are the Challenges and Opportunities of Using Text AI in the DHH Community?}

We identify the challenges the DHH community faces when using Text AI, followed by suggestions and opportunities to address these challenges and improve the tools' accessibility and inclusivity.

\subsubsection{Challenge 1: Crafting Effective Prompts in English} 

Among participants that have previously used Text AI, 44.1\% reported difficulties in crafting effective prompts, 23.5\% experienced technical issues (such as network connectivity issues), 22.1\% noted limited sign language support, 14.7\% expressed doubts about AI's utility, and 10.3\% found responses difficult to understand. The primary challenge was shaping questions in English, akin to prompt engineering, to achieve specific or accurate results. As one participant (S53) noted, \textit{``As a native ASL user, it takes time to generate the right prompt because it's text-only.''} This highlights the effort required for ASL users to translate goals into effective prompts, especially when conveying nuanced or complex text input. 

\textbf{Suggestion: Text AI Should Support ASL and Diverse Language Needs} 
Most Text AI tools rely on moderate-to-high English proficiency of users, limiting accessibility for users with lower proficiency. Participants recommended features like ASL video explanations using virtual avatars to enhance accessibility and provide tailored feedback for prompt improvements. While ASL recognition and generation through avatars remain imperfect \cite{naert2020survey, desai2024asl, kezar2023sem, starner2024popsign}, a practical first step could involve training Text AI on ASL gloss. More generally, participants highlighted the need for Text AI to better understand and improve prompts written in ASL gloss. As S22 suggested, \textit{``Maybe another option is to ask the LLM to respond in ASL grammar order.''} Training Text AI to recognize ASL gloss would require specialized datasets sourced from the DHH community and guided by fairness and disability-first principles \cite{theodorou2021disability}. This adaptation could help the tool \textit{``better learn to understand prompts from Deaf users''} (S2), making it more inclusive and effective, and serve as a valuable language-learning aid, reducing the cognitive burden of translating between ASL and English.

\textbf{Suggestion: DHH-Focused Tutorials and Instructional Design for Text AI} 
Participants stressed the importance of DHH-friendly instructions to address language deprivation and teach effective use of Text AI. Suggestions included offering classes tailored to DHH users in schools and colleges (S64) and creating resources explaining \textit{``how Text AI can help particularly Deaf/HoH users''} (S50). %Such tutorials could encourage broader adoption, address misconceptions, and empower the DHH community to use Text AI confidently. 
Developing such tutorials requires significant resources and expertise in both Text AI and the DHH community. While institutions like higher education have created tutorials for instructors and students \cite{jin2024generative}, similar efforts for the DHH community are lacking and should be prioritized. When top-down approaches fall short, bottom-up methods, like crowdsourcing, may empower underrepresented groups to lead efforts and share knowledge  \cite{inzerillo2016crowdsourcing} and should be studied in future research.

\subsubsection{Challenge 2: Perceived Shallow Understanding of The DHH community by Text AI}
Participants felt that Text AI tools lack a nuanced understanding of the Deaf community’s diversity, where deafness may be viewed as anything from a medical condition to a cultural identity. This gap limits the AI’s ability to address important historical and social contexts. For example, S79 compared ChatGPT and Copilot's explanations of the word \textit{``dumb''} and noted that while ChatGPT provided historical context about its relevance to disability, Copilot omitted this information\footnote{A comparison
of the chat histories is included in the appendix.}. Although the term ``deaf and dumb'' is now considered outdated and inappropriate, it remains part of Deaf history, and its historical context should still be shared as a fact when such terms arise. Participants also criticized Text AI’s adherence to standardized notions of \textit{``professionalism,''} which they perceived as rooted in \textit{``white hearing culture.''} For example, S44 remarked that the overly formal phrasing, such as \textit{``I hope this email finds you well,''} felt inauthentic and disconnected from their natural communication style. This disconnect highlights a need for AI to account for diverse cultural norms and communication preferences.

\textbf{Suggestion: Incorporating Deaf Community Input into AI Models}
Beyond ASL support, participants emphasized the need for Text AI to integrate Deaf cultural knowledge and linguistic nuances through training on data from Deaf individuals. As S44 noted, \textit{``It needs further representation of culturally Deaf ideas, practices, and communication in the data sets.''} Incorporating cultural awareness would help Text AI interpret terms like \textit{``dumb''} accurately and foster more inclusive interactions. 
%For example, understanding the relationship between linguistic development and cultural identity could inform model training. Research shows that DHH reading comprehension is shaped by phonological and language awareness during key developmental phases \cite{easterbrooks2015reading}. Incorporating such insights into dataset selection and model design would equip Text AI to better meet the unique communication needs of DHH individuals, enhancing both accessibility and inclusivity.
%LLMs may not fully understand the Deaf community. 
Further research is needed to explore how to incorporate Deaf community input and address discrepancies between LLM understanding and real-world experiences. Additionally, integrating existing research, such as how DHH reading comprehension varies with phonological and language awareness during key developmental phases \cite{easterbrooks2015reading}, is essential for improving personalized interactions for individual users.

\textbf{Suggestion: More Research on How and Why DHH Individuals Disclose to Text AI}
A few participants shared cultural, linguistic, and accessibility details with Text AI, finding it helpful for refining prompts. For example, S76 explained, \textit{``In ChatGPT it has a memory, I have it remembered I’m an expert user of English and ASL.''} None of our participants intentionally avoided discussing their English proficiency or background when using Text AI. Our findings may be skewed towards individuals who see deafness as a cultural identity rather than a medical condition, with 68.8\% self-identifying as Deaf. Moreover, research shows that identity evolves over time; for instance, DHH adolescents often integrate other significant aspects, such as ethnicity and culture, into their context-dependent d/Deaf identity \cite{byatt2021evolving}. Future research should explore how DHH individuals self-identify with AI and whether they adapt identity cues to receive tailored responses.

\subsection{Limitations} 
Despite ASL translation and video submission options, the survey was not fully accessible to all participants. One participant requested a video call, and video uploads posed challenges for some \cite{mack2020social}. Accessibility for DeafBlind users was limited, with only one participant despite multiple invitees; ProTactile\footnote{Protactile is a tactile language used by the DeafBlind community \cite{morrison2017protactile, granda_nuccio2018protactile}.} options could have improved inclusion. The survey was primarily distributed within the Deaf community, potentially underrepresenting Hard of Hearing individuals, and university-based recruitment skewed the sample toward college students. Future research should broaden recruitment to include a more diverse DHH population and those less affiliated with Deaf culture.

%% file: 5-conclusion.tex
\section{Conclusion}
This study explored DHH individuals' experiences with Text AI tools, highlighting both benefits and barriers. Participants used Text AI to improve communication with hearing individuals through English support and cultural bridging, alleviating much of the effort DHH people typically invest to facilitate communication. While Text AI helps reduce gaps between Deaf and hearing communities, challenges persist, including difficulties with English prompts, limited understanding of Deaf culture, and the need for Deaf-specific instructional materials. Participants called for ASL input/output support, better adaptation to ASL grammar, and AI trained on Deaf culture. These findings provide insights into the unique ways the DHH community interacts with Text AI and offer recommendations to improve these tools for accessibility and inclusion. Future research should focus on developing AI that better serves underserved and underrepresented communities.

%% file: 7-appendix.tex
%TC:ignore
\section{Text AI Tools’ Responses to the Prompt “Dumb meaning?”}

\begin{table}[htbp]
    %\caption{Text AI’s responses to the prompt “Dumb meaning?”}
    \label{tab:dumb_meaning}
    \centering
    \renewcommand{\arraystretch}{1.2} 
    \begin{tabular}{p{0.65\textwidth}|p{0.3\textwidth}}
        \hline
        \textbf{ChatGPT's Response} & \textbf{Copilot's Response} \\ 
        \hline
        The word "dumb" has several meanings, depending on context:\newline\newline
        1. \textbf{Lacking intelligence:} Informally, "dumb" is often used to describe someone or something as unintelligent or foolish. For example, "That was a dumb decision."\newline
        2. \textbf{Unable to speak:} Historically, "dumb" was used to describe someone who is unable to speak, often due to a disability. This usage is now considered outdated and offensive, with "nonverbal" or "mute" being more appropriate terms.\newline
        3. \textbf{Silenced or quiet:} In some contexts, "dumb" can refer to being silent or not speaking. For example, "struck dumb" means being rendered speechless, often due to shock or surprise.\newline\newline
        It's important to use the word carefully, as calling someone "dumb" can be hurtful or offensive. 
        & Silly, foolish, not exactly the sharpest tool in the shed. What's driving your curiosity about that word? \\
        \hline
    \end{tabular}
\end{table}

% \begin{figure}[ht]
%     \centering  
%     \includegraphics[width=0.7\textwidth]{figures/screenshot_gpt.png}    
%     \caption{ChatGPT's response to a query about the meaning of `dumb.'} 
%     \Description{Screenshot showing a participant asking ChatGPT, `dumb meaning?' ChatGPT responds with several meanings: 1. Lacking intelligence, used informally to describe someone or something as unintelligent or foolish. 2. Historically, used to describe someone unable to speak due to a disability, but this usage is now considered outdated and offensive, with `nonverbal' or `mute' as more appropriate terms. 3. Refers to being silent or speechless, as in `struck dumb.' ChatGPT emphasizes using the word carefully, as it can be hurtful or offensive.}
%     \label{fig:gpt}
% \end{figure}

% \begin{figure}[ht]
%     \centering  
%     \includegraphics[width=0.7\textwidth]{figures/screenshot_copilot.png}
%     \caption{Copilot's response to a query about the meaning of `dumb.'}
%     \Description{Screenshot showing a participant asking Copilot, `dumb meaning?' Copilot responds: `Silly, foolish, not exactly the sharpest tool in the shed. What's driving your curiosity about that word?'}
%     \label{fig:copilot}
% \end{figure}\textbf{}

\section{Survey Questions}

Definition: 
\textbf{Text AI}, short for text-based generative AI, is a type of computer program that can read and write human languages. Users interact with Text AI by inputting text prompts, to which the AI responds with generated text. Common examples are OpenAI’s ChatGPT, Microsoft’s Copilot, and Google’s Gemini.

\textbf{Section 1: Usage of Text AI}

1. How often do you use Text AI (please see Text AI’s definition above)?
\begin{enumerate}
    \item Daily
    \item Several times a week
    \item Rarely
    \item Never (redirect to \textbf{Section 6}, and then \textbf{Section 5})
\end{enumerate}

2. Which Text AI  do you usually use? (Select all that apply)
\begin{enumerate}
    \item OpenAI ChatGPT
    \item Microsoft Copilot
    \item Google Gemini
    \item Meta Llama
    \item Other (please specify)
\end{enumerate}

3. For what purposes do you use Text AI? You can use either ASL or English to answer the question. (Open ended)

4. If you use Text AI for English reading or writing related tasks, what are they? (Select all that apply)
\begin{enumerate}
    \item Writing emails
    \item Text messaging
    \item Social media posts
    \item Work-related communication
    \item School-related communication
    \item Other (please specify)
    \item I do not use Text AI for English-related tasks
\end{enumerate}

\textbf{Section 2: English Proficiency Improvement}

1. What is your comfort level using written English to communicate \textbf{without} using Text AI?
\begin{enumerate}
    \item Very uncomfortable
    \item Somewhat uncomfortable
    \item Fair
    \item Comfortable
    \item Very comfortable
\end{enumerate}

2. What is your comfort level using written English to communicate \textbf{with} Text AI?
\begin{enumerate}
    \item Very uncomfortable
    \item Somewhat uncomfortable
    \item Fair
    \item Comfortable
    \item Very comfortable
\end{enumerate}

3. When you use Text AI, do you specify your English level in your initial chat with the Text AI (aka prompt)? For example, “English is not my first language so please simplify your answer”. You can use either ASL or English to answer the question. (Open ended)

4. Other than the “trick” above, do you have other “tricks” that you use to make the Text AI  generate something more suitable for you? You can use either ASL or English to answer the question. (Open ended)

5. Which aspects of your English writing have improved (if any) while using Text AI? (Select all that apply)
\begin{enumerate}
    \item Topics and ideas
    \item Organization and logical order
    \item Tone and voice (conveying the writer’s feelings, attitude, and emotion)
    \item Vocabulary and word choice
    \item Sentence fluency and readability
    \item Conventions (rule of spelling, punctuation, capitalization, and grammar)
    \item Overall confidence in writing
    \item Other (please specify)
\end{enumerate}

\textbf{Section 3: Communication Barriers and Discrimination}

1. \textbf{Without} using Text AI, how often do you feel anxious / nervous when communicating with hearing people in writing?
\begin{enumerate}
    \item Daily
    \item Several times a week
    \item Rarely
    \item Never
\end{enumerate}

2. \textbf{With} Text AI, how often do you feel anxious / nervous when communicating with hearing individuals in writing?
\begin{enumerate}
    \item Daily
    \item Several times a week
    \item Rarely
    \item Never
\end{enumerate}

3. Do you feel that you have experienced increased or reduced discrimination since using Text AI?
\begin{enumerate}
    \item Significantly decreased (much less discrimination)
    \item Somewhat decreased (less discrimination)
    \item No change
    \item It has increased (more discrimination)
    \item I can’t tell
\end{enumerate}

\textbf{Section 4: Satisfaction and Suggestions}

1. How satisfied are you with the assistance provided by Text AI in improving your English proficiency?
\begin{enumerate}
    \item Very satisfied
    \item Satisfied
    \item Neutral
    \item Dissatisfied
    \item Very dissatisfied
\end{enumerate}

2. Following up on your answer above: why or why not are you very satisfied / satisfied / neutral / dissatisfied / very satisfied with the assistance provided by Text AI? You can use either ASL or English. (Open ended)

3. What difficulties have you encountered while using Text AI? (Select all that apply)
\begin{enumerate}
    \item Difficulty understanding responses
    \item Limited sign language support
    \item Difficulty in asking questions / finding the right prompts
    \item Technical issues, such as network connectivity issues
    \item Not sure how Text AI can help
    \item Other (please specify)
\end{enumerate}

4. Do you have any suggestions for improving Text AI assistance for Deaf/HoH users? You can either use ASL or English. (Open ended)

\textbf{Section 5: Demographic Information}

1. What is your age?
\begin{enumerate}
    \item 18-25
    \item 26-35
    \item 36-50
    \item 51 and above
\end{enumerate}

2. What is your highest level of education?
\begin{enumerate}
    \item High school / GED or less
    \item Some college
    \item Associate degree or equivalent
    \item Bachelor’s degree
    \item Graduate degree
    \item Other (please specify)
\end{enumerate}

3. What is your identity in the DHH community? 
\begin{enumerate}
    \item Deaf
    \item deaf
    \item Hard of hearing
    \item DeafBlind
    \item Other (please specify)
\end{enumerate}

4. What is your first language? (Select all that apply)
\begin{enumerate}
    \item English
    \item ASL
    \item Other (please specify)
\end{enumerate}

5. What is your primary language for communication?
\begin{enumerate}
    \item English
    \item ASL
    \item Other (please specify)
\end{enumerate}

6. We are thinking of running interview sessions with a subset of this survey’s respondents. Would you like to be contacted about this opportunity?
\begin{enumerate}
    \item Yes
    \item No
\end{enumerate}

7. What's your email address?

\textbf{Section 6: Possibility of using Text AI (for people who have never used it)}

1. What are the main reasons you have not used Text AI? (Select all that apply)
\begin{enumerate}
    \item Not aware of text AI tools
    \item Privacy concerns
    \item Lack of trust in AI
    \item Difficulty in using technology
    \item Prefer human assistance
    \item Not sure what Text AI can do
    \item Other (please specify)
\end{enumerate}

2. Would you consider using Text AI in the future?
\begin{enumerate}
    \item Yes
    \item No
    \item Unsure
\end{enumerate}

3. If you were to consider using Text AI, what tasks do you think you might use them for? (Select all that apply, and please think beyond  AI’s current abilities)
\begin{enumerate}
    \item Improving English writing skills
    \item Translating ASL to English
    \item Translating English to ASL
    \item Learning new vocabulary
    \item Practicing English grammar
    \item Reducing anxiety before communicating with hearing people
    \item Other (please specify)
\end{enumerate}

4. What features or improvements would make you more likely to use Text AI? You can either use ASL or English. (Open ended)

%TC:endignore